\documentclass{article}

 \usepackage{amsmath}
 \usepackage{amsfonts}

\newtheorem{theorem}{Theorem}

\input{tcilatex}

\begin{document}

\title{Corrugated surfaces with slow modulation and quasiclassical
Weierstrass representation}

\author{B.G.Konopelchenko \\
%$^{\ast }$
\\ Dipartimento di Fisica, Universita di Lecce
\\and INFN, Sezione di Lecce, 73100 Lecce, Italy}
\date{}
\maketitle

\begin{abstract}

\bigskip

Quasiclassical generalized Weierstrass representation for highly corrugated
surfaces in $\mathbb{R}^{3}$ with slow modulation is proposed. Integrable
deformations of such surfaces are described by the dispersionless modified
Veselov-Novikov hierarchy.

\end{abstract}

\section{Introduction}
\setcounter{equation}{0}

Surfaces, interfaces, fronts and their dynamics are the key ingredients in a
number of very interesting phenomena from hydrodynamics, growth of crystals,
theory of membranes to the string theory and gravity (see \textit{e.g.} [1-5]).
The most of the papers on this subject and results obtained have been concerned
to a smooth case. On the other hand, irregular, corrugated surfaces also
have attracted interest in various fields from the applied physics,
technology to the pure mathematics (see \textit{e.g} [6-13]).

In the present paper we propose a Weierstrass type representation for highly
corrugated surfaces with a slow modulation in the three dimensional
Euclidean space $\mathbb{R}^{3}$. \ It is the quasiclassical limit of the generalized
Weierstrass representation (GWR) for surfaces in $\mathbb{R}^{3}$ introduced in
[14,15]. This GWR was based on the two-dimensional Dirac equation and it
allows to construct any surface in $\mathbb{R}^{3}$.
The hierarchy of the modified Veselov-Novikov (mVN) equations provides us
with the integrable deformations of surfaces [14,15].

The quasiclassical GWR is based on the quasiclassical limit of the Dirac
equation. It allows us to construct surfaces in $\mathbb{R}^{3}$ with highly
oscillating (corrugated) profiles and slow modulations of these oscilations
characterized by a small parameter $\varepsilon =\frac{l}{L}$ where l and L
are typical scales of oscillations and modulations, respectively. In the
lowest order in $\varepsilon $ the coordinates $X^{j}$ ($j=1,2,3$) of such
surfaces are of the form
\begin{eqnarray}
X^{1}+iX^{2} &=&A(\varepsilon z,\varepsilon \overline{z})
\exp \left[-2i\frac{S(\varepsilon z,\varepsilon \overline{z})}{\varepsilon } \right],
\nonumber
\\
X^{3} &=&\frac{1}{\varepsilon }B(\varepsilon z,\varepsilon \overline{z})
\label{1.1}
\end{eqnarray}
where $z$ and $\overline{z}$ are the conformal coordinates on a surface, A
and B some smooth functions and S is a solution of the eikonal equation. The
corresponding metric and mean curvature are finite functions of the slow
variables $\varepsilon z,\varepsilon \overline{z}$ while the Gaussian
curvature is of the order $\varepsilon ^{2}$.

Integrable deformations of such corrugated surfaces are induced by the
hierarchy of dispersionless mVN equations . These deformations preserve the
quasiclassical Willmore functional (Canham-Helfrich bending energy for
membranes or the Polyakov extrinsic action for strings (see \textit{e.g.}
[3,4]). The quasiclassical limit of the Gauss map and
Kenmotsu representation
for surfaces in $\mathbb{R}^{3}$ of above type are discussed too.

\section{Generalized Weierstrass representation for surfaces in
$\mathbb{R}^{3}$}
\setcounter{equation}{0}

The generalized Weierstrass representation (GWR) proposed in [14.15] is
based on the linear system ( two-dimensional Dirac equation)
\begin{eqnarray}
\Psi _{z} &=&p\Phi ,  \nonumber \\
\Phi _{\overline{z}} &=&-p\Psi
\end{eqnarray}
where $\Psi $ and $\Phi $ are complex -valued functions of $z,\overline{z}%
\in \mathbb{C}$ (bar denotes a complex conjugation ) and $p(z,\overline{%
z})$ is a real-valued function. One defines three real-valued functions $%
X^{j}(z,\overline{z}),j=1,2,3$ by the formulae
\begin{eqnarray}
X^{1}+iX^{2} &=&i\int_{\Gamma }(\overline{\Psi }^{2}dz\prime -\overline{\Phi
}^{2}d\overline{z\prime }), \nonumber \\
X^{1}-iX^{2} &=&i\int_{\Gamma }(\Phi ^{2}dz\prime -\Psi ^{2}d\overline{%
z\prime }), \\
X^{3} &=&-\int_{\Gamma }(\overline{\Psi }\Phi dz\prime +\Psi \overline{\Phi }%
d\overline{z\prime }) \nonumber
\end{eqnarray}
where $\Gamma $ is an arbitrary contour in $\mathbb{C}$.

\begin{theorem} {\bf [14,15]}.
For any function $\ p(z,\overline{z})$ and any solution ($%
\Psi ,\Phi $) of the system (2.1) the formulae (2.2) define a conformal
immersion of a surface into $\mathbb{R}^{3}$ with the induced metric
\begin{equation}
ds^{2}=u^{2}dzd\overline{z,}
\end{equation}
the Gaussian curvature
\begin{equation}
K=-\frac{4}{u^{2}}(\log u)_{z\overline{z}},
\end{equation}
mean curvature
\begin{equation}
H=2\frac{p}{u},
\end{equation}
and the Willmore functional given by
\begin{equation}
W\overset{def}{=}\iint_{G}W^{2}[ds]=4\iint_{G}p^{2}dxdy,
\end{equation}
where $u=\left| \Psi \right| ^{2}+\left| \Phi \right| ^{2}$ and $z=x+iy$.
\end{theorem}

Morever, any regular surface in $\mathbb{R}^{3}$
can be constructed via the GWR (2.1),(2.2).

Integrable dynamics of surfaces constructed via the GWR (2.2) is induced by
the integrable evolutions of the potential $p(z,\overline{z},t)$ and the
functions $\Psi ,\Phi $ with respect the \ deformation parameter t. \ They
are given by the modified Veselov-Novikov (mVN) hierarchy [14,15]. The
simplest example is the mVN equation

\begin{eqnarray}
p_{t}+p_{zzz}+p_{\overline{z}\overline{z}\overline{z}}+3\omega p_{z}+3%
\overline{\omega }p_{\overline{z}}+\frac{3}{2}p\omega _{z}+\frac{3}{2}p%
\overline{\omega }_{\overline{z}} &=&0,\nonumber \\
\omega _{\overline{z}} &=&(p^{2})_{z}
\end{eqnarray}
while $\Psi $ and $\Phi $ obey the system of linear equations
\begin{eqnarray}
\Psi _{t}+\Psi _{zzz}+\Psi _{\overline{z}\overline{z}\overline{z}%
}-3p_{z}\Phi _{z}+3\overline{\omega }\Psi _{\overline{z}}+\frac{3}{2}%
\overline{\omega }_{\overline{z}}\Psi +3p\omega \Phi &=&0, \nonumber \\
\Phi _{t}+\Phi _{zzz}+\Phi _{\overline{z}\overline{z}\overline{z}}+3\omega
\Phi _{z}+3p_{\overline{z}}\Psi _{\overline{z}}-3p\overline{\omega }\Psi +%
\frac{3}{2}\omega _{z}\Phi &=&0.
\end{eqnarray}

The mVN equation (2.7) and the whole mVN hierarchy are amenable to the
inverse spectral transform method [16,15] and they have a number of
remarkable properties typical for integrable 2+1-dimensional equations.
Integrable dynamics of surfaces in $\mathbb{R}^{3}$ inherits all these
properties [15]. One of the remarkable features of such dynamics is that the
Willmore functional W (2.6) remains invariant ($W_{t}=0$) [17,18]. In virtue
of the linearity of the basic problem (2.1) the GWR is quite a useful tool
to study various problems in physics and mathematics (see \textit{e.g.}
[18-21]).

A different representation of surfaces in $\mathbb{R}^{3}$ has been
proposed in [22]. It is based on the following parametrisation of the Gauss
map
\begin{equation}
\overrightarrow{G}=(1-f^{2},i(1+f^{2}),2f),
\end{equation}
where $f(z,\overline{z})$ is a complex-valued function. Then the Kenmotsu
representation of a surface is given by [22]
\begin{equation}
\overrightarrow{X}(z,\overline{z})=\func{Re}\left( \int_{\Gamma }\eta
\overrightarrow{G}dz\prime \right) ,
\end{equation}
where the function $\eta $ obeys the compatibility condition
\begin{equation}
\left( \log \eta \right) _{\overline{z}}=-\frac{2\overline{f}f_{\overline{z}}%
}{1+\left| f\right| ^{2}},
\end{equation}

The Kenmotsu representation (2.9)-(2.11) and the GWR (2.1), (2.2) are
equivalent to each other. The relation between the functions $(f,\eta )$ \
and $(\Psi ,\Phi )$ is the following [17,18]
\begin{equation}
f=i\frac{\overline{\Psi }}{\Phi },\eta =i\Phi ^{2},
\end{equation}
while
\begin{equation*}
p=-\frac{\eta f_{\overline{z}}}{\left| \eta \right| (1+\left| f\right| ^{2})}.
\end{equation*}

Both the GWR and Kenmotsu representations have been widely used to study
properties of generic surfaces and special classes of surfaces, in
particular, of the constant mean curvature surfaces.

\section{Quasiclassical Weierstrass representation.}
\setcounter{equation}{0}

In this paper we shall consider a class of surfaces in $\mathbb{R}^{3}$
which can be characterized by two scales l and L such that the parameter
$\varepsilon =\frac{l}{L}\ll 1.$
A simple example of such a surface is provided by the
profile of a slowly modulated wavetrain for which l is a typical wavelength
and L is a typical length of modulation. Theory of such highly oscillating
waves with slow modulations is well developed (see \textit{e.g.} [1,23]).
Borrowing
the ideas of this Whitham (or nonlinear WKB) theory we study surfaces in
$\mathbb{R}^{3}$ for which the coordinates $X^{1},X^{2},X^{3}$ have the form

\begin{equation}
X^{i}(z,\overline{z})=\sum_{n=0}^{\infty }\varepsilon ^{n}F_{n}^{i}\left(
\frac{\overrightarrow{S}(\varepsilon z,\varepsilon \overline{z})}{%
\varepsilon },\varepsilon z,\varepsilon \overline{z}\right) , \quad i=1,2,3
\end{equation}
where $\overrightarrow{S}=(S^{1},S^{2},S^{3})$ and $F_{n}^{i}$ are smooth
functions of slow variables $\xi =\varepsilon z,\overline{\xi }=\varepsilon
\overline{z}$ and the small parameter $\varepsilon $ is defined above. The
arguments $\frac{S^{i}}{\varepsilon }$ in $F_{n}^{i}$ describe a relatively
fast variation of a surface while the rest of arguments correspond to slow
modulations.

There are different ways to specify functions $\ F_{n}^{i}$. Here we will
consider one of them induced by the similar quasiclassical (WKB) limit of
the GWR (2.1), (2.2).

Thus, we begin with the quasiclassical limit of the Dirac equation (2.1).
Having in mind the discussion given above, we take

\begin{equation*}
p=\sum_{n=0}^{\infty }\varepsilon ^{n}p_{n}(\varepsilon z,\varepsilon
\overline{z}),
\end{equation*}

\begin{equation}
\Psi =\exp \left( \frac{iS(\varepsilon z,\varepsilon \overline{z})}{%
\varepsilon }\right) \sum_{n=0}^{\infty }\varepsilon ^{n}\Psi
_{n}(\varepsilon z,\varepsilon \overline{z}),
\end{equation}

\begin{equation*}
\Phi =\exp \left( \frac{iS(\varepsilon z,\varepsilon \overline{z})}{%
\varepsilon }\right) \sum_{n=0}^{\infty }\varepsilon ^{n}\Phi
_{n}(\varepsilon z,\varepsilon \overline{z})
\end{equation*}
where $S,\Psi _{n},\Phi _{n}$ are smooth functions of slow variables $\xi
=\varepsilon z,\overline{\xi }=\varepsilon \overline{z}$ and $\overline{S}=S$%
. \ Substituting (3.2) into (2.1), one in zero order in $\varepsilon $ gets

\begin{eqnarray}
iS_{\xi }\Psi _{0}-p_{0}\Phi _{0} &=&0, \nonumber \\
p_{0}\Psi _{0}+iS_{\overline{\xi }}\Phi _{0} &=&0
\end{eqnarray}
while in the order $\varepsilon $ one has
\begin{eqnarray}
iS_{\xi }\Psi _{1}-p_{0}\Phi _{1} &=&-\Psi _{0\xi }+p_{1}\Phi _{0,} \nonumber \\
p_{0}\Psi _{1}+iS_{\overline{\xi }}\Phi _{1} &=&-\Phi _{0\overline{\xi }%
}-p_{1}\Psi _{0}.
\end{eqnarray}

The existence of nontrivial solutions for the system (3.3) implies that S
should obey the equation%
\begin{equation}
\det \left|
\begin{array}{cc}
iS_{\xi } & -p_{0} \\
p_{0} & iS_{\overline{\xi }}%
\end{array}%
\right| =-S_{\xi }S_{\overline{\xi }}+p_{0}^{2}=0,
\end{equation}
or $(\xi =\xi _{1}+i\xi _{2})$
\begin{equation}
S_{\xi _{1}}^{2}+S_{\xi _{2}}^{2}=4p_{0}^{2},
\end{equation}
\textit{i.e.} the Hamilton-Jacobi equation for the two-dimensional classical system
with the potential $4p_{0}^{2}$  or the eikonal equation with the refraction
index $4p_{0}^{2}$ .

From equations (3.3) and (3.5) it follows that $\left| S_{\xi }\right|
=\left| S_{\overline{\xi }}\right| =p_{0}$ and $\left| \Psi _{0}\right|
=\left| \Phi _{0}\right|$. Equations (3.4) imply the $iS_{\overline{\xi }%
}(\Psi _{0\xi }-p_{1}\Phi _{0})=p_{0}(\Phi _{0\overline{\xi }}+p_{1}\Psi
_{0}).$

Using the differential form of the formulae (2.2), \textit{i.e.}
\begin{equation}
\left( X^{1}+iX^{2}\right) _{z}=i\overline{\Psi }^{2},\left(
X^{1}+iX^{2}\right) _{\overline{z}}=-i\overline{\Phi }^{2},X_{z}^{3}=-%
\overline{\Psi }\Phi ,
\end{equation}
one concludes that the coordinates $X^{i}$ have the form
\begin{eqnarray}
X^{1}+iX^{2} &=&\sum_{n=0}^{\infty }\varepsilon ^{n}A_{n}(\varepsilon
z,\varepsilon \overline{z})\exp \left( -\frac{2iS(\varepsilon z,\varepsilon
\overline{z})}{\varepsilon }\right) , \nonumber \\
X^{3} &=&\frac{1}{\varepsilon }\sum_{n=0}^{\infty }\varepsilon
^{n}B_{n}(\varepsilon z,\varepsilon \overline{z})
\end{eqnarray}
where $A_{n}$ and $B_{n}$ are smooth functions. In the lowest order in $%
\varepsilon $ one has
\begin{equation}
A_{0}=-\frac{\overline{\Psi }_{0}^{2}}{2S_{\xi }}=\frac{\overline{\Phi }%
_{0}^{2}}{2S_{\overline{\xi }}},B_{0\xi }=-\overline{\Psi} _{0}\Phi _{0}.
\end{equation}
Note that both the expressions for $A_{0}$ are equivalent to each other due
to equations (3.3) and (3.5).

Thus, we have the

\begin{theorem}. The quasiclassical GWR provides us with the highly
corrugated (oscillating) surfaces with slow modulations for which the
coordinates have the form (3.8) where the function S is a solution of the
eikonal equation (3.5) and in the lowest order the functions $A_{0}$ and $%
B_{0}$ are given by (3.9),(3.3). In the limit $\varepsilon \rightarrow 0$
one has the following principal contributions to the metric
\begin{equation}
ds^{2}=4\left| \Psi _{0}(\varepsilon z,\varepsilon \overline{z})\right|
^{4}dzd\overline{z}=\frac{4}{\varepsilon ^{2}}\left| \Psi (\xi ,
\overline{\xi })\right| ^{4}d\xi d\overline{\xi ,}
\end{equation}
mean curvature
\begin{equation}
H_{0}(\xi ,\overline{\xi })=\frac{p_{0}(\xi ,\overline{\xi })}{2\left| \Psi
_{0}(\xi ,\overline{\xi }\right| ^{2}},
\end{equation}
the Gaussian curvature
\begin{equation}
K_{0}=-\varepsilon ^{2}\frac{2}{\left| \Psi _{0}(\xi ,\overline{\xi }\right|
^{2}}\left( \log \left| \Psi _{0}\right| \right) _{\xi \overline{\xi }%
},
\end{equation}
and the Willmore functional
\begin{equation}
W_{0}=4\iint_{G}p_{0}^{2}(\varepsilon z,\varepsilon \overline{z})dxdy=\frac{4%
}{\varepsilon ^{2}}\iint_{G_{\varepsilon }}p_{0}^{2}(\xi ,\overline{\xi }%
)d\xi _{1}d\xi _{2},
\end{equation}
where $G_{\varepsilon }$ is the rescaled domain $G(x=\frac{\xi _{1}}{%
\varepsilon },y=\frac{\xi _{2}}{\varepsilon })$.
\end{theorem}

One can refer to surfaces in $\mathbb{R}^{3}$ given by the formulae (3.8)-(3.13) as
the quasiclassical surfaces. They represent a subclass of surfaces of the
type (3.1).

We note that, in virtue of (3.6), the Willmore functional $W_{0}$ is just
the Dirichlet integral
\begin{equation}
W_{0}=\frac{1}{\varepsilon ^{2}}\iint_{G_{\varepsilon }}\left( S_{\xi
_{1}}^{2}+S_{\xi _{2}}^{2}\right) d\xi _{1}d\xi _{2}.
\end{equation}

So, due to the Dirichlet principle (see \textit{e.g.} [24]) the problem of
minimization of $W_{0}$ is equivalent to the Dirichlet boundary problem for
the harmonic function in the domain $G_{\varepsilon }$. For surfaces with
all $p_{n}=0,n=1,2,3...$ . the formulae (3.13) and (3.14) give us not just
the asymptotic expressions for the Willmore functional at $\varepsilon
\rightarrow 0$ , but the exact one.

The quasiclassical analogs of surfaces of constant mean curvature and
surfaces with $H\sqrt{\det g}=1$ (see \textit{e.g.} [18]) correspond to the
constraints $p_{0}=2\left| \Psi _{0}\right| ^{2}$ and $4p_{0}\left| \Psi
_{0}\right| ^{2}=1$ , respectively. In the very particular case $p_{1}=0$
and $\Psi _{1}=\Phi _{1}=0$ one has $\Psi _{0\xi }=0,\Phi _{0\overline{\xi }%
}=0$, \textit{i.e.}
$\Psi _{0}=\Psi _{0}(\overline{\xi }),\Phi _{0}=\Phi _{0}(\xi )$.
Consequently, the formulae (3.3), (3.8)-(3.13) generate developable surfaces
($K_{0}=0$) with the metric $ds_{0}^{2}=\frac{4}{\varepsilon ^{2}}\Psi
_{0}^{2}(\overline{\xi })\overline{\Psi }_{0}^{2}(\xi )d\xi d\overline{\xi }$
which after the reparametization $d\xi \rightarrow dw=2\overline{\Psi }%
_{0}^{2}(\xi )d\xi $ becomes $ds_{0}^{2}=\frac{1}{\varepsilon ^{2}}dwd%
\overline{w\text{.}}$

Now let us consider the quasiclassical limit of the Kenmotsu representation.
In virtue of (2.12) one has

\begin{eqnarray}
f &=&\widetilde{f}\exp \left( -\frac{2iS(\varepsilon z,\varepsilon \overline{%
z})}{\varepsilon }\right) , \nonumber \\
\eta &=&\widetilde{\eta }\exp \left( \frac{2iS(\varepsilon z,\varepsilon
\overline{z})}{\varepsilon }\right)
\end{eqnarray}
where
\begin{eqnarray*}
\widetilde{f} &=&\sum_{n=0}\varepsilon ^{n}f_{n}(\varepsilon z,\varepsilon
\overline{z}), \\
\widetilde{\eta } &=&\sum_{n=0}\varepsilon ^{n}\eta _{n}(\varepsilon
z,\varepsilon \overline{z})
\end{eqnarray*}
and $f_{0}=i\frac{\overline{\Psi _{0}}}{\Phi _{0}},h_{0}=i\Phi _{0}^{2}$ and
so on. Using (3.15), one gets the quasiclassical Gauss map
\begin{equation*}
\overrightarrow{G_{q}}=\left( 1-\widetilde{f}^{2}e^{-\frac{4iS}{\varepsilon }%
},i(1+\widetilde{f}^{2}e^{-\frac{4iS}{\varepsilon }}),2\widetilde{f}e^{-%
\frac{2iS}{\varepsilon }}\right)
\end{equation*}
and, finally, the quasiclassical Kenmotsu representation
\begin{equation}
\overrightarrow{X}=\func{Re}\left\{ \frac{1}{\varepsilon }\int_{\Gamma
_{\varepsilon }}d\xi \prime \widetilde{f}(\xi \prime ,\overline{\xi }\prime )%
\widetilde{\eta }(\xi \prime ,\overline{\xi }\prime )\left( \widetilde{f}%
^{-1}e^{\frac{2iS}{\varepsilon }}-\widetilde{f}e^{-\frac{2iS}{\varepsilon }%
},i(\widetilde{f}^{-1}e^{\frac{2iS}{\varepsilon }}+\widetilde{f}e^{-\frac{2iS%
}{\varepsilon }}),2\right) \right\} ,
\end{equation}
which , of course, is equivalent to the quasiclassical GWR (3.8).

Quasiclasical versions of the GWR's for surfaces in the 4-dimensional
[25,26,21] and higher dimensional spaces can be constructed in a similar
manner.

\section{Integrable deformations via the dmVN hierarchy.}
\setcounter{equation}{0}

Deformations of quasiclassical surfaces described above are given by the
dispersionless limit of the mVN hierarchy. To get this limit one, as usual
(at \ the 1+1-dimensional case, see \textit{e.g.} [27]), assumes that the dependence
of all quantities on t is a slow one, \textit{i.e.}
$p=p(\varepsilon z,\varepsilon
\overline{z},\varepsilon t),S=S(\varepsilon z,\varepsilon \overline{z}%
,\varepsilon t)$ and so on. At the first and second orders in $\varepsilon $
equation (2.7) gives ($\tau =\varepsilon t,\omega =\sum_{n=0}\varepsilon
^{n}\omega _{n}(\varepsilon z,\varepsilon \overline{z},\varepsilon t))$
\begin{eqnarray}
p_{0\tau }+3\omega _{0}p_{0\xi }+3\overline{\omega _{0}}p_{0\overline{\xi }}+%
\frac{3}{2}p_{0}\omega _{0\xi }+\frac{3}{2}p_{0}\overline{\omega }_{0%
\overline{\xi }} &=&0, \nonumber \\
\omega _{0\overline{\xi }} &=&(p_{0}^{2})_{\xi }
\end{eqnarray}
and
\begin{eqnarray}
p_{1\tau }+3\omega _{1}p_{0\xi }+3\omega _{0}p_{1\xi }+3\overline{\omega }%
_{0}p_{1\overline{\xi }}+3\overline{\omega }_{1}p_{0\overline{\xi }}+\qquad \qquad  &&
\nonumber \\ \qquad \qquad \qquad +\frac{3%
}{2}p_{1}(\omega _{0\xi }+\overline{\omega }_{0\overline{\xi }})+\frac{3}{2}%
p_{0}(\omega _{1\xi }+\overline{\omega }_{1\overline{\xi }}) &=&0, \nonumber \\
\omega _{1\overline{\xi }} &=&2(p_{0}p_{1})_{\xi }.
\end{eqnarray}

In the lowest order equation (2.8) is
\begin{equation}
M_{0}\left(
\begin{array}{c}
\Psi _{0} \\
\Phi _{0}%
\end{array}%
\right) =0,
\end{equation}
where
\begin{equation}
M_{0}=\left(
\begin{array}{cc}
i(S_{\tau }-S_{\xi }^{3}-S_{\overline{\xi }}^{3}+3\overline{\omega }_{0}S_{%
\overline{\xi }}), & 3p_{0}\omega _{0} \\
-3p_{0}\overline{\omega }_{0}, & i(S_{\tau }-S_{\xi }^{3}-S_{\overline{\xi }%
}^{3}+3\omega _{0}S_{\xi })%
\end{array}%
\right)
\end{equation}

With the use of (3.5) the condition $\det M_{0}=0$  assumes the form

\begin{equation}
\left( S_{\tau }-S_{\xi }^{3}-S_{\overline{\xi }}^{3}\right) \left( S_{\tau
}-S_{\xi }^{3}-S_{\overline{\xi }}^{3}+3\omega _{0}S_{\xi }+3\overline{%
\omega }_{0}S_{\overline{\xi }}\right) =0,
\end{equation}

Equation (4.1) is the dispersionless limit of the mVN equation (dmVN
equation). It is equivalent to the compatibility condition of the linear
systems (3.3) and (4.3). In a similar manner one constructs the whole dmVN
hierarchy.

This hierarchy generates the integrable deformations of the quasiclassical
surfaces described in the previous section via the $\tau $- dependence of $%
p_{0},p_{1,}\Psi _{0},\Phi _{0}$ ect  given by equations (4.1)-(4.5) and so
on. These integrable deformations are very special from the geometrical
viewpoint. Indeed, the dmVN equation (4.1) implies that

\begin{equation}
\left( p_{o}^{2}\right) _{\tau }+3\left( \omega _{0}p_{0}^{2}\right) _{\xi
}+3\left( \overline{\omega }_{0}p_{0}^{2}\right) _{\overline{\xi }}=0.
\end{equation}

So, for periodic or rapidly decaying at $\left| \xi \right| \rightarrow
\infty $ functions $p_{0}$ and $\omega _{0}$ one has

\begin{equation}
W_{0\tau }=\frac{4}{\varepsilon ^{2}}\iint_{G_{\varepsilon }}\left(
p_{0}^{2}\right) _{\tau }d\xi _{1}d\xi _{2}=0.
\end{equation}

One can show that the Willmore functional $W_{0}$ (or Dirichlet integral
(3.14)) is invariant under the whole dmVN hierarchy of deformations as well.
One may suggest that the quasiclassical limit of the higher mVN integrals ,
discussed in [28], will provide us with the higher geometrical invariants
for quasiclassical surfaces.

The formula (4.6) indicates also an interesting connection of the dmVN
equation with the other known dispersionless equation. Indeed, denoting
$p_{0}^{2}=u$, one has
\begin{eqnarray}
u_{\tau }+3\left( \omega _{0}u\right) _{\xi }+3\left( \overline{\omega }%
_{0}u\right) _{\overline{\xi }} &=&0, \nonumber \\
\omega _{0\overline{\xi }} &=&u_{\xi .}
\end{eqnarray}

It is the dispersionless VN equation introduced in [29,30]. The dVN equation
is equivalent to the compatibility condition for the two Hamilton-Jacobi
equations
\begin{eqnarray}
S_{\xi }S_{\overline{\xi }}-u &=&0, \nonumber\\
S_{\tau }-S_{\xi }^{3}-S_{\overline{\xi }}^{3}+3\omega _{0}S_{\xi }+3%
\overline{\omega }_{0}S_{\overline{\xi }} &=&0.
\end{eqnarray}

These equations show that the whole theory of the dVN hierarchy can be
developed without any reference to its dispersive version. Within the
quasiclassical $\overline{\partial }-$ dressing \ method the dVN hierarchy
has been studied in [31] in connection with the problems of nonlinear
geometrical optics in the so-called Cole-Cole media. \ In order to apply the
results obtained for the dVN hierarchy to the dmVN hierarchy one, due to the
relation $u=p_{0}^{2}$ , should be able to select effectively the positive
solutions of the dVN hierarchy. This problem and also the application of the
\ $\overline{\partial }-$ dressing method directly to the dmVN hierarchy
will be considered elsewhere.

The quasiclassical $\overline{\partial }-$ dressing method is based on the
nonlinear Beltrami equation for the function S in the auxuliary space of
''spectral'' parameter . This approach reveals a deep interrelation between
the solutions of the eikonal equation (3.6) and \ the dVN hierarchy and the
quasiconformal mappings on the plane [31]. One may suggest that the theory
of the quasiclassical surfaces presented in the section 3 is closely related
to the theory of quasiconformal mappings on the plane too.

\bigskip

\textbf{Acknowledgements. } This work is supported in part by COFIN
PRIN '' Sintesi'' 2004.

\bigskip

\section*{References}

\begin{enumerate}
\item G.B. Whitham, Linear and Nonlinear waves, John Wiley \&Sons, New York,
1974.

\item A.R.Bishop, L.J.Campbell and P.J.Channell, Eds., Fronts, Interfaces and
Patterns, North-Holland, 1984.

\item A.M.Polyakov, Gauge Fields and Strings, Harwood Acad.Publ., Chur,1987.

\item D.Nelson, T.Piran abd S.Weinberg, Eds., Statistical Mechanics of
Membranes and Surfaces, World Scientific, Singapore, 1989.

\item F.David, P.Ginsparg and Y.Zinn-Justin, Eds., Fluctuating Geometries in
Statistical Mechanics and Field Theory, Elsevier, Amsterdam, 1996.

\item E.Feinberg, On the propagation of radio waves along an imperfect surface,
Acad. Sci. USSR. J. Phys., \textbf{8} (1944), 317-330.

\item R.A.Hurd, The propagation of an electromagnetic wave along an infinite
corrugated surface, Canad. J. Phys., \textbf{32 }(1954), 727-734.

\item S.Asano, Reflection and refraction of elastic waves at a corrugated
boundary surface, Bull. Earthquake Res. Inst. Tokyo, I,\textbf{\ 38} (1960),
177-197: II, \textbf{39 }(1961), 367-466.

\item I.H.Sabitov, The rigidity of corrugated surfaces of revolution, Mat.
Zametki, \textbf{34 }(1973), 517-522.

\item J.Krug and P.Meakin, Kinetic roughening of Laplacian fronts, Phys. Rev.
Lett., \textbf{66 }(1991), 703-706.

\item D.A.Lidar, Inversion of randomly corrugated surface structure from atom
scattering data, Inverse Problems, \textbf{14 }(1998), 1299-1310.

\item V.Jaksic, Spectral theory of corrugated surfaces, Journees'' Equations
aux Derives Part. '', \textbf{VIII }\ (2001), 11.

\item Y.Tsori and \ D.Andelman, Parallel and perpendicular lamellae on
corrugated surfaces, Macromolecules, \textbf{36 }(2003), 8560-8573. \ \ \

\item B.G.Konopelchenko, Multidimensional integrable systems and dynamics of
surfaces in space ( preprint of Institute of Mathematics, Taipei, Taiwan,
August 1993), in Proc. '' National Workshop on Nonlinear Dynamics'' (
M.Costato, A.Degasperis and M.Milani, Eds.), pp.33-40, Ital. Phys. Soc.,
Bologna, 1995.

\item B.G.Konopelchenko, Induced surfaces and their integrable dynamics, Stud.
Appl. Math., \textbf{96 }\ (1996), 9-51.

\item L.V.Bogdanov, The Veselov-Novikov equation as a natural two-dimensional
generalisation of the Korteweg- de Vries equation, Teor Mat. Fiz., \textbf{%
70 }(1987), 309-317.

\item B.G.Konopelchenko and I. Taimanov, Generalized Weierstrass formulae,
soliton equations and Willmore surfaces, preprint N.187, Univ. Bochum., 1995.

\item R.Carroll and B.G.Konopelchenko, Generalized Weierstrass-Enneper
inducing, conformal immersion and gravity, Int. J. Mod. Phys. A, \textbf{11}
(1996), 1183-1216.

\item I.Taimanov, Modified Novikov-Veselov equation and differential geometry,
Trans. Amer. Math. Soc., Ser.2, \textbf{179 }(1997), 133-159.

\item B.G.Konopelchenko and G.Landolfi, Quantum effects for extrinsic geometry
of strings via the generalized Weierstrass representation,
Phys. Lett. B, \textbf{444 }(1998), 299-308.

\item F.Pedit and U.Pinkall, Quaternionic analysis on Riemann surfaces and
differential geometry, \ Doc. Math.-Extra volume ICM 1998, II (1998),
389-400.

\item K.Kenmotsu, Weierstrass formulae for surfaces of prescribed mean
curvature, Math. Ann., \textbf{245 }(1979), 89-99.

\item B.Dubrovin and S.Novikov, Hydrodynamics of weakly \ deformed soliton
lattices: differential geometry and Hamiltonian theory, Russian Math.
Surveys, \textbf{44 }(1989), 35-124.

\item A. Hurwitz and R.Courant, Funktionentheorie, Springer-Verlag, Berlin,
1964.

\item B.G.Konopelchenko, Weierstrass representation for surfaces in 4D spaces
and their integrable deformations via DS hierarchy, Annals of Global Anal.
and Geom., \textbf{16 }(2000), 61-74; arXiv; math.DG/9807129 (1998).

\item B.G.Konopelchenko and G.Landolfi, Induced surfaces and their integrable
dynamics. II. Generalized Weierstrass representation in 4D spaces and
deformations via DS hierarchy, Stud. Appl. Math., \textbf{104 }(1999),
129-169.

\item V.E.Zakharov,  Benney equations and quasiclassical approximation in the
method of the inverse problem, Funk. Anal. Pril., \textbf{14 }(1980), 89-98.

\item P.G.Grinevich and M.V.Schmidt, Conformal invariant functionals of
immersion of tori into $R^{3}$, J. Geom. Phys., \textbf{26 }(1998), 51-78.

\item I.M.Krichever, Averaging method for two-dimensional integrable
equations, Func. Anal. Pril., \textbf{22 }(1988), 37-52.

\item B.G.Konopelchenko and L.Martinez Alonso, Nonlinear dynamics on the plane
and integrable hierarchies of infinitesimal deformations, Stud. Appl. Math.,
\textbf{109 }(2002), 313-336.

\item B. Konopelchenko and A. Moro, Integrable equations in nonlinear
geometrical optics, Stud. Appl. Math., \textbf{\ 113 }(2004) 325-352.

\end{enumerate}

\end{document}